# Controlling Strain Bursts and Avalanches at the Nano-to-Micro Scale

Yinan Cui,[*] Giacomo Po, and Nasr Ghoniem

*Mechanical and Aerospace Engineering Department, University of California,
Los Angeles, 420 Westwood Plaza, Los Angeles CA, 90095*

We demonstrate, through 3-dimensional discrete dislocation dynamics simulations, that the complex dynamical response of nano and micro crystals to external constraints can be tuned. Under load rate control, strain bursts are shown to exhibit scale-free avalanche statistics, similar to critical phenomena in many physical systems. For the other extreme of displacement rate control, strain burst response transitions to quasi-periodic oscillations, similar to stick-slip earthquakes. External load mode control is shown to enable a qualitative transition in the complex collective dynamics of dislocations from self-organized criticality to quasi-periodic oscillations.

Power-law scaling of avalanche phenomena is widely observed in many nonequilibrium natural systems. Examples are found in geologic earthquakes, snow avalanches, sand pile slides, and strain bursts during plastic flow [1, 2]. The realization that such vastly diverse physical systems display common features, implies scale invariance and compels a search into universal fundamental laws. The common scaling raises the possibility that the intricate system behavior can be described by simple local rules, despite the complexity of the underlying internal dynamics. One concept that is widely used to interpret this universality is self-organized criticality (SOC) [3]. In a SOC system, the dynamics has an attractor characterized by infinite correlation time and length, hence displaying scale-free scaling. A key hypothesis behind this abstraction is that the driving force varying rate is much slower than the internal relaxation rate [3, 4] of a system undergoing SOC. Nevertheless, since this condition may not always hold, one wonders if the qualitative aspects of a system's dynamical behavior change when the driving force changing rate is comparable to its internal relaxation rate? Our objective here is to investigate the relationship between the external driving force and relaxation dynamics associated with strain bursts during nano- and micro-scale plastic deformation of crystals.

At the smallest of physical scales (e.g. nano-to-micro scale), the release of plastic strain by intermittent "bursts" has been found to belong to this power-law scaling behavior [2, 5–8]. One additionally unique aspect of plasticity is that the driving force varying rate can be experimentally tailored. Considering a simple but illustrative case, a pillar is subjected to uniaxial compression in Fig. 1. The force actuator, typically a voice coil, can exert an open-loop stress rate $\dot{\sigma}_0$ and/or be controlled to impose a strain rate $\dot{\varepsilon}_0$. For a proportional controller with stiffness $K_p$, the internal stress rate in the pillar is [9],

$$\dot{\sigma} = \frac{\alpha E}{1+\alpha}(\dot{\varepsilon}_0 - \dot{\varepsilon}^p) + \frac{\dot{\sigma}_0}{1+\alpha} \qquad (1)$$

where $\alpha = K_p/K$ is the relative stiffness ratio, $K = EA/H$ is the pillar stiffness, $E$, $A$ and $H$ are the Young module, cross section area and height of the pillar, respectively. $\dot{\varepsilon}^p$ is the plastic strain rate due to all internal dislocation dynamical activities. Once the stiffness ratio $\alpha$ is infinitely large, or $\dot{\sigma}_0$ and $\dot{\varepsilon}_0$ are very low, $\dot{\sigma}$ becomes very sensitive to $\dot{\varepsilon}^p$, implying that the driving force changing rate ($\dot{\sigma}$) is dominated by and comparable to its internal relaxation rate ($\dot{\varepsilon}^p$). This indicates that the corresponding slip statistics are expected to violate SOC.

However, it is generally believed that the machine stiffness $K_p$ only contributes to the cutoff of the power law scaling [6, 8, 10]. The present investigation demonstrates that, if the machine stiffness is extremely high, dislocation avalanche dynamics (and hence strain bursts) undergo a transition from scale-free critical behavior to quasi-periodic oscillations. Interestingly, this is consistent with recent findings on the role of very slow loading rates (low $\dot{\sigma}_0$ and $\dot{\varepsilon}_0$) [11, 12], as suggested by Eq. 1. The underlying microstructure mechanism for this dynamical regime transition are disclosed. Considering that the dynamical behaviors under soft or hard machine stiffness conditions are vastly different, the corresponding intermittent plasticity will henceforth be described as either *avalanche* or *burst*, respectively. Moreover, a dislocation based branching model is proposed, giving a clear and precise physical picture of the avalanche dynamical behavior.

The vast majority of existing submicron mechanical testing experiments can only cover a narrow range of machine stiffness. In addition, the time necessary for dislocations to travel through 1 $\mu$m sample is estimated at about 1 ns [13]. In state-of-the-art experiments, the feedback loop frequency is $\approx$ 78 kHz (time constant $\approx$ 13 $\mu$s) [8], which means that current experimental controller response rate is much slower than sample plastic relaxation rate by 4 orders of magnitude. Namely, the driving force changing rate is much slower than internal relaxation rate. Therefore, most previous experimental conditions correspond to the regime where SOC is observed. Discrete dislocation dynamics (DDD) studies, as a computer simulation tool, make it possible to supple-ment experimental testing and explore regimes that are currently difficult to access experimentally [6, 14]. The current research presents the first systematic 3D-DDD investigation on the slip statistics at submicron scale, accounting for the effects of the interaction of an external loading mode [15–17]. Compared with most of existing two dimensional (2D) DDD studies [2, 18], the key approximations inherent in 2D techniques are resolved. Specifically, dislocation junction formation and destruction, and the occurrence of cross slip are all accounted for with minimal ad hoc assumptions.

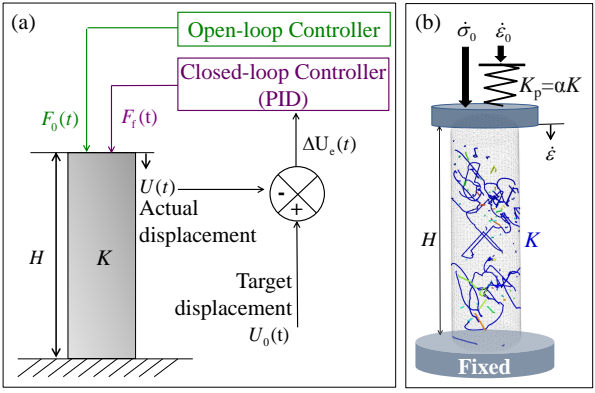

FIG. 1. Simplified sketch of pillar compression. (a) Experimental setup with an open-loop (directly applying a force $F_0$) and closed-loop control (to realize displacement control); (b) Simulation setup, a proportional dominated closed-loop control is considered here with $F_f = K_p(U_0 - U)$, which is simplified as a spring with a finite machine stiffness $K_p$. The external stress rate $\dot{\sigma}_0 = \dot{F}_0/A$, target strain rate $\dot{\varepsilon}_0 = \dot{U}_0/H$, actual strain rate $\dot{\varepsilon} = \dot{U}/H$, where $A$ and $H$ are the cross section area and height of the pillar, respectively. One typical dislocation configuration in a pillar with $d$ =3000 b is shown as an example

The simulation setup is schematically shown in Fig. 1b. We conducted simulations of compression tests on Cu pillars of different diameters, ranging from 1000-3000 b ($\approx$ 300 nm- 1 $\mu$m), where $b$ is the burgers vector magnitude. The aspect ratio H/d is 3. Two extreme machine stiffness cases are first considered, corresponding to *pure strain control* ($\alpha = +\infty$) and *pure stress control* ($\alpha = 0$). Here, under pure stain control, the applied strain rate $\dot{\varepsilon}_0 = 960 s^{-1}$. Correspondingly, under pure stress control, the actual loading rate $\dot{\sigma}_0$ is $E\dot{\varepsilon}_0$. Fifty and twenty separate simulations with different initial dislocation configurations are carried out under each loading mode for $d$ =1000 b and $d$ =3000 b, respectively.

Figure 2a presents the results of statistical analysis of the burst displacement magnitude $\Delta U$. To obtain maximum resolution of the limited simulation data set, the complementary cumulative distribution function (CCDF) is used. Fig. 2a clearly illustrates that $\Delta U$, under pure stress control, exhibits a well-defined power law distribution spanning several orders of magnitude. The power law exponent for the corresponding probability density is found to be 1.5, agreeing well with the generally accepted range of 1.35 $\sim$ 1.67 [5, 6, 19–21]. In addition, the power law distribution is consistent across system size, implying the existence of scale-free universality. In contrast, the CCDF of $\Delta U$ under pure strain control seems not to exhibit power-law scaling behavior for both small and large system sizes. Meanwhile, most of the data concentrate within one order of magnitude. An analogous breakdown of the power law scaling under pure strain control is also observed for the statistics of burst duration [9].

Then, how to describe the strain burst statistics under pure strain control? When discussing the temporal statistics of earthquakes, distinct dynamical behaviors are distinguished by the coefficient of variation $C = s_x/\overline{x}$ [22], where $s_x$ and $\overline{x}$ are the standard deviation and mean value, respectively. For the cases of $C > 1$ and $C < 1$, the distribution is refereed to as "clustered" and "quasi-periodic", respectively; otherwise, if $C = 1$, it is a random Poisson distribution [22]. Taking the results of $\Delta U$ here, $C$ is calculated as 1.9 and 0.9 under pure stress and pure strain control, respectively. This suggests that the dynamical behaviors under pure strain control becomes quasi-periodic. Similar to previous studies [11, 22], quasi-periodicity here is found to be stochastic, due to the intrinsic scatter induced by random cross slip or different dislocation configurations. Quasi-periodic strain bursts under pure strain control are manifested through the smoothed plastic strain rate, as clearly shown in Fig. 2b. Here, the time series of $\dot{\varepsilon}^p$ is smoothed over a fixed time window of 0.24 $\mu$s. For comparison, the smoothed plastic strain rate under pure stress control, also shown in Fig. 2c, corresponds to a depinning phase transition.

Close examination of dislocation configuration evolution reveals that the mechanisms that control avalanche versus quasi-periodic burst behavior are significantly different, and are highly dependent on the external constraint. First, let's consider pure strain control. In the submicron regime (e.g. $d$ =1000 b), each strain burst is found to be dominated by sequential activation and deactivation of single arm dislocation sources. Once a source is activated, the accompanying plastic strain leads to a decrease in the stress level (see Eq. 1, $\alpha = +\infty$). Even if a weaker source is formed during one burst event, sometimes it also cannot operate due to the lower prevailing stress after relaxation. This makes it difficult to trigger simultaneous operation of multiple dislocation sources (see Fig. 3b), especially for small samples with limited volume. We have recently shown that dislocation sources themselves are transient, because they generally result from the formation of dipolar loops by cross-slip [7]. This rapid stress drop prevents the strain burst from continuously growing into a full-fledged avalanche. Consequently, large-scale cooperative interactions between dislocations that can lead to SOC cannot be realized under pure strain control. Note that this discussion applies to a sample size ranging from several nanometers to about 1 micrometer. For smaller pillars, surface nucleation of dislocations becomes dominant [23], and the rapid stress drop may inhibit correlated surface nucleation, while for larger pillar size, Taylor-type interaction mechanisms prevail [24, 25], and the rapid stress drop may suppress cooperative dislocation interactions.

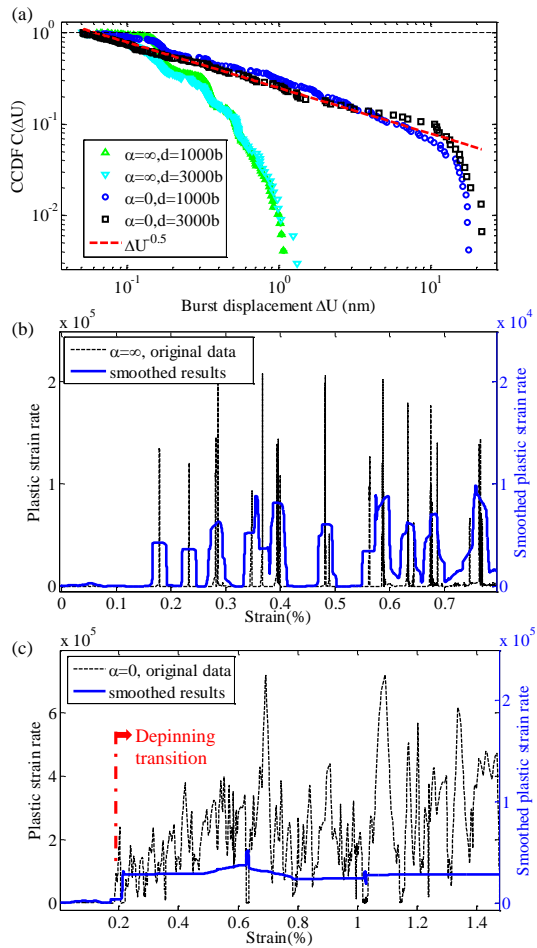

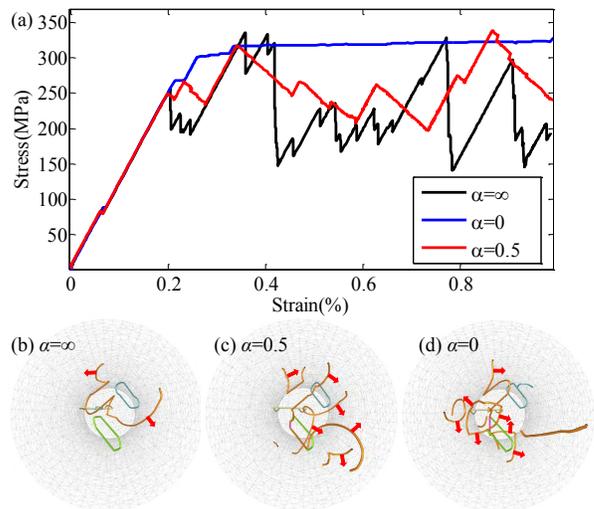

FIG. 2. (a) Statistical properties of burst displacement under pure strain and stress control modes for pillar with diameters $d = 1000$ b and 3000 b. (b-c) Typical evolution of plastic strain rate and its averaged value in 0.24 $\mu$s windows, showing (b) quasi-periodic strain bursts under pure strain control, and (c) depinning transition dislocation avalanche under pure stress control

FIG. 3. Typical simulation results under different loading modes for pillar with $d$ =1000 b. (a) Stress-strain curves; (b-d) Snapshots of dislocation configurations (from top view) at a strain value of 0.4%, arrows indicate the bowing out directions of activated sources

By contrast, dislocation avalanche under pure stress control is clearly associated with correlated dislocation motion. According to Eq. 1, when $\alpha = 0$, the stress rate cannot sense the internal dislocation activity. Thus, the stress level keeps almost constant during each avalanche event (see Fig. 3a). If one activated source leads to the formation of a weaker one, it can be immediately activated. Thus, distinctly different from the strain control case discussed above, multiple sources can operate in a correlated fashion (see Fig. 3d). All correlated sources contribute then to an increasing magnitude of the strain burst, turning it into an "avalanche". Such highly correlated dynamical behavior suggests a close-to-criticality nonequilibrium state [3].

Since it is difficult to experimentally achieve such extreme machine stiffness, it is then interesting to examine dislocation dynamics with finite machine stiffness. All the results in Fig. 3a correspond to the same size and initial dislocation configuration. The calculated stress-strain curve with finite machine stiffness ($\alpha = 0.5, \dot{\sigma}_0 = 0$) in Fig. 3a displays a very similar behavior to experimental results [8, 21], and exhibits a serrated yield character with longer decaying stages as compared to pure strain control. The observation of simultaneous operation of multiple sources in Fig. 3c suggests that a finite machine stiffness actually promotes correlated dislocation motion, compared with pure strain control.

To further elucidate the statistical difference between avalanche versus quasi-periodic dynamics, a simple dislocation based branching model is proposed. It is inspired by the present 3D-DDD simulations, and motivated by Zapperi's sand-pile branching model [26], in which we translate the branching idea into dislocation language. The discrete plastic deformation is assumed to mainly proceed through the intermittent activation of dislocation sources [27, 28]. One activated source may lead to the stochastic generation/activation of other sources, similar to a branching process shown in Fig. 4a.

The detailed algorithm proceeds as follows. Assuming a pillar initially with $n_s$ dislocation sources, we can randomly give each source a specific length $\lambda$ according to a given source length probability distribution. The fate of each source (active or not) is determined by checking whether the instantaneous applied stress $\sigma_k$ can reach the source operation stress,

$$\sigma_k \cdot M \geqslant \tau_0 + \alpha_1 \mu b \sqrt{\rho} + \alpha_2 \mu b / \lambda \qquad (2)$$

where $M$ is Schmid factor, the three terms on the right hand are lattice friction stress, the elastic interaction stress described by Taylor relation, and the source strength, respectively. $\alpha_1$ and $\alpha_2$ are dimensionless constant, set to 0.5 and 1 [28], respectively. $\rho$ is the instantaneous dislocation density, estimated by dividing the total source length by the pillar volume.

Once the weakest source is activated during deformation, a strain burst begins [28, 29]. After each source is activated, the burst strain $S_k$ increases by a specific value $d\varepsilon^p$. Considering that $\dot{\varepsilon}^p$ is much higher than the applied strain rate $\dot{\varepsilon}_0$ during a strain burst, according to Eq. 1, $\sigma_k$ drops by $Ed\varepsilon^p \alpha/(\alpha+1)$, and the total strain increases by $d\varepsilon^p/(\alpha+1)$. It is assumed that the activated source is broken (ceases to operate) after it sweeps the entire slip plane once. However, it can randomly trigger the generation of additional $n_a$ sources. If the newly generated source can be activated according to Eq. 2, it triggers subsequent generation of $n_a$ sources. Otherwise, the new source is stored for possible dislocation generation, which may activate during subsequent deformation stages. This branching source generation process repeats itself until all dislocation sources cannot be activated under the combined effect of the instantaneous applied stress and the resistance stress, given by the right side of Eq. 2 (see Fig. 4a). At this instance, this strain burst event stops and the stress continues to increase till it triggers another strain burst event.

In the following, we investigate the slip statistics using this abstract branching model, and compare to the more fundamental DDD simulations discussed above. Compression tests are also modeled for Cu pillars with diameter $d$=1000 b and 3000 b. Similar to DDD simulation, surface nucleation is not considered. If the stress is higher than the surface nucleation stress (about 1.2 GPa for Cu [30]) or if the strain is higher than 0.5, events are not recorded. If there is only one activated source, each burst strain corresponds to the generated plastic strain when the dislocation sweeps the entire slip plane once. Therefore, $d\varepsilon^p$ is set to $bM/H/cos\beta$ [28], where $\beta$ is the angle between the normal direction of the slip plane and the loading orientation. Through examination of the dislocation configuration evolution, $n_a$ is taken as the nearest integer of $2 \cdot rand$, where $rand$ represents a random value from 0 to 1. Accordingly, the probabilities of $n_a$ being 0, 1 and 2 are 25%, 50% and 25%, respectively. This is different from previous sand-pile branch-ing model [26], where the new activated site number was taken a constant value of 2. $n_a = 0$ means that the source is destroyed after operation once, $n_a = 1, 2$ indicate that other sources are generated due to interactions with other dislocations, cross slip, forming superjogs, or forming dipolar loops [7]. Note that, more deactivated sources may be left in the sample if $n_a$=2, leading to a slight increase in the dislocation density $\rho$ after each branching process. This results in an increase in the elastic interaction resistance stress. Similar to 2D-DDD simulations [31], the source length is assumed to follow a Gaussian distribution, with a mean value $\overline{\lambda} = d/2$, determined according to the yield stress of our DDD results. Its standard deviation is set to $20\%\overline{\lambda}$, so that the predicted activated source number for each strain burst event is statistically equivalent to those obtained by our DDD results under pure strain control (see Fig. 4c).

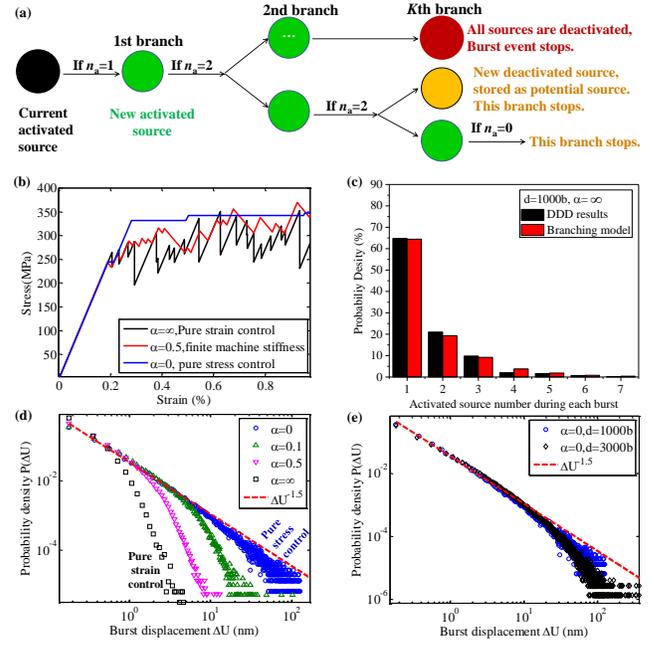

FIG. 4. (a) Schematic showing the random branching dislocation source generation and activation process, $n_a$ is the number of newly generated dislocation sources, green filled circles represent that new source is activated, only activated source may trigger further branching process; (b-d) Typical predicted results for pillar with $d$ =1000 b, (b) Stress-strain curve, (c) Comparison of activated source number during each burst under pure strain control, (d) Probability density function of burst displacement for different machine stiffness; (e) Probability density function of burst displacement for different sample sizes

Fig. 4b presents predicted typical stress-strain curves under different loading modes, which agree well with our simulation results in Fig. 3a, including the stress level and the stepped or serrated burst features. In addition, the power law scaling of burst displacement $\Delta U$ is also well reproduced under pure stress control for different pillar sizes in Fig. 4e. The power law exponent of the probability distribution of $\Delta U$ agrees with that obtained by the present 3D-DDD. Fig. 4d clearly indicates that as the machine stiffness increases, the power law tails gradually become too wide to recognize proper scale-free power law statistics.

The excellent agreement between the abstract branching model prediction and the fundamental 3D-DDD simulations further verify that hard machine stiffness leads to deviation from scale-free SOC, because the rapid stress relaxation disturbs correlated dislocation motion. The current finding offers a new pathway towards controlling the correlated extent of dislocation dynamics and the intermittent statistics by tuning the machine stiffness. It opens up new possibilities for novel experiments with faster response rate that can reveal the quasi-periodic oscillation dynamics of dislocation systems. The importance of often-neglected interaction with the external loading system on intermittent plastic flow has been demonstrated. The complex dynamics of collective dislocations producing strain bursts is shown to be controlled through simple tuning of the relative value of driving force rate to internal relaxation rate.

This material is based upon work supported by the U.S. Department of Energy, Office of Science, Office of Fusion Energy Sciences, under Award Number DE-FG02-03ER54708, and the US Air Force Office of Scientific Research (AFOSR), under award number FA9550-11-1-0282. We would like to thank Professor Michael Zaiser (Friedrich-Alexander University Erlangen-Nuremberg) and Professor Stephanos Papanikolaou (Johns Hopkins University) for inspiring comments and discussions.